\documentstyle[12pt,fleqn]{article}
\def\theequation{\arabic{section}.\arabic{equation}}
\def\thesection {\Roman{section}.}
\def\thesubsection {\Alph{subsection}.}
\renewcommand{\section}[1]{\pagebreak[3]\par\vspace{0.6em}
\addtocounter{section}{1}\begin{center}{\bf \thesection\ #1}
\end{center}\vspace{-0.8ex}\setcounter{equation}{0}
\setcounter{subsection}{0}}
\renewcommand{\subsection}[1]{\pagebreak[3]\par\vspace{0.4em}
\addtocounter{subsection}{1}\begin{center}{\bf \thesubsection\ #1}
\end{center}\vspace{-0.4ex}}
\renewcommand{\appendix}{\renewcommand{\thesection}
{APPENDIX}\renewcommand{\theequation}{\Alph{section}\arabic{equation}}
\setcounter{equation}{0}\setcounter{section}{0}}

\newcounter{eqncmplx}

\renewcommand{\cite}[1]{\ref{#1}}
\renewcommand{\bibitem}[1]{\item\label{#1}}
\renewenvironment{thebibliography}{\pagebreak[3]\par\vspace{0.6em}
\begin{center}{\bf References}\end{center}\vspace{-1.0em}

\begin{enumerate}\if@twocolumn\baselineskip=0.6em\itemsep -0.2em
\else\itemsep -0.2em\fi\labelsep 0.1em}{\end{enumerate}}

\newenvironment{figurecaptions}{\pagebreak[3]\par\vspace{0.6em}
\begin{center}{\bf Figure Captions}\end{center}\vspace{-1.0em}
\newcounter{fig}\begin{list}{$\;$ FIG.\ \arabic{fig}.}{
\usecounter{fig}\setlength{\rightmargin}{\leftmargin}}}{\end{list}}

\def\abstract{\begin{center}{\bf Abstract}\end{center}\quotation}

\newcommand{\BQ}{\begin{equation}}
\newcommand{\EQ}{\end{equation}}
\newcommand{\BQA}{\begin{eqnarray}}
\newcommand{\EQA}{\end{eqnarray}}
\newcommand{\half}{\frac{1}{2}}
\newcommand{\NN}{\nonumber\\}
\newcommand{\e}{{\rm e}}
\newcommand{{\BPsi}}{\bar\Psi}
\newcommand{{\bpsi}}{\bar\psi}
\newcommand{\GF}{\gamma_5}
\newcommand{\Gmu}{\gamma^\mu}

\newcommand{\gmu}{\gamma_\mu}

\newcommand{\Z}{{\bf Z}}

\newcommand{\dagg}{\mbox{\scriptsize{\dag}}}
\newcommand{\dagpsi}{\psi^{\mbox{\scriptsize{\dag}}}}
\newcommand{\ket}[1]{\vert #1 \rangle}             
\newcommand{\bra}[1]{\langle #1 \vert}             
\newcommand{\ketrm}[1]{\vert {\rm #1} \rangle}     
\newcommand{\RG}{\cR[\cG]}
\newcommand{\RGC}{\cR[\cG_{\rm C}]}
\newfont{\cmsy}{cmsy10 at 12pt}
\renewcommand{\cal}{\cmsy}
\newcommand{\cL}{\mbox{\cal L}}
\newcommand{\cH}{\mbox{\cal H}}
\newcommand{\cG}{\mbox{\cal G}}
\newcommand{\cJ}{\mbox{\cal J}}
\newcommand{\cR}{\mbox{\cal R}}
\newcommand{\reg}{{\rm reg}}
\newcommand{\Hcl}{H_{\rm cl}}
\newcommand{\PR}{Phys.\ Rev.\ }
\newcommand{\PRL}{Phys.\ Rev.\ Lett.\ }
\newcommand{\PL}{Phys.\ Lett.\ }
\newcommand{\NP}{Nucl.\ Phys.\ }
\newcommand{\AP}{Ann.\ Phys.\ }

\newcommand{\PTP}{Prog.\ Theor.\ Phys.\ }

\begin{document}

\setlength{\topmargin}{0cm}
\setlength{\oddsidemargin}{0cm}
\setlength{\baselineskip}{6mm}

\begin{titlepage}
\vspace{1cm}
\vspace*{-2cm}
\begin{flushright}UT-Komaba 94-5\end{flushright}
\vspace{2cm}
\begin{center}
{\LARGE Quantum Hamiltonian Reduction\\
\vspace{5mm}
of the Schwinger Model}\\

\vspace{2cm}

{\large Kazunori I{\sc takura}\footnote{e-mail address:
itakura@hep1.c.u-tokyo.ac.jp
} and Koichi O{\sc hta}}\\

\vspace{1.0cm}

{\large\it Institute of Physics, University of Tokyo,\\ Komaba,
Meguro-ku, Tokyo 153, Japan }\\

\vspace{2.5cm}

{\bf Abstract}
\end{center}
We reexamine a unitary-transformation method of extracting a physical
Hamiltonian from a gauge field theory after quantizing all degrees
of freedom including redundant variables.  We show that this {\it quantum
Hamiltonian reduction} method suffers from crucial modifications
arising from regularization of composite operators. We assess the
effects of regularization in the simplest gauge field theory, the
Schwinger model. Without regularization, the quantum reduction yields
the identical Hamiltonian with the classically reduced one. On the
other hand, with regularization incorporated, the resulting
Hamiltonian of the quantum reduction disagrees with that of the
classical reduction. However, we find that the discrepancy is
resolved by redefinitions of fermion currents and that the
results are again consistent with those of the classical reduction.

\end{titlepage}

\newpage

\section{INTRODUCTION}

Gauge theories are constructed to have physically irrelevant
variables and therefore gauge fields need to be constrained to
avoid having too many degrees of freedom.
 Identification of physical degrees of freedom is indispensable
 for extracting physical information from them.
Although this identification is not difficult in Abelian gauge
theories, we know that it is a highly non-trivial problem in
non-Abelian gauge theories  such as quantum chromodynamics (QCD):
The gauge-fixing ambiguities prevent us
 from constructing the theory nonperturbatively.

 The Hamiltonian formulation of a constrained
system was founded by Dirac [\cite{Dirac}]. In his method, all
variables including unphysical ones are treated on the equal
footing. Constraints of a first-class system,
$\Phi^j=0$, commute with each other weakly in the Poisson brackets,
(i.e., they commute on the constraint surface) and generate gauge
transformations. Therefore a first-class constraint system is
always a gauge system. Gauge degrees of freedom are eliminated by
imposing gauge-fixing conditions so that the system becomes second
class, whereby the Poisson brackets between constraints do not vanish
any more. For a second-class system, we define Dirac's brackets and
replace them with the commutators to quantize the system.
If the system is a gauge theory, we can also proceed as follows:
Ignoring  constraints,  we first quantize all variables except for
Lagrange multipliers, and then reduce the extended Hilbert space to
the physical subspace by imposing
\BQ
\Phi^j\ketrm{phys}=0.\label{genecon}
\EQ
In this canonical Weyl- or temporal-gauge formulation,
quantization precedes reduction of the Hilbert space.
In this point, this is different from schemes such as the
Faddeev-Jackiw formalism [\cite{FadJac},\cite{Jac}], in which
quantization is performed only for physical variables in the
classically reduced phase space.  Various arguments have been made
about the equivalence of these two approaches, for which there seems
to be no a priori justification (see [\cite{Loll},\cite{OP}] and
references therein). Furthermore, since the Hamiltonian in the
canonical temporal-gauge formulation explicitly contains
unphysical variables and constraints are complicated in general, it
is hard to calculate various physical quantities.

Recently, Lenz et al.\ [\cite{LNOT}] have revamped a
unitary-transformation method that is based on the canonical
temporal-gauge formulation and gives without
gauge-fixing procedure a {\it physical Hamiltonian}, i.e., a Hamiltonian
described only by physical variables [\cite{GerSak}-\cite{UtiSaka}].
We shall call this method the quantum Hamiltonian reduction (QHR).
This method makes it easy to identify unphysical variables.
By performing  a suitable unitary transformation $U$,
we can eliminate unphysical coordinates from the Hamiltonian,
\BQ
UHU^{\dagg}=H'(q^i,p^i,P^j),
\EQ
where $q^i$ and $p^i$  are physical coordinates and their conjugate
momenta, and $P^j$ are unphysical momenta, respectively.
Simultaneously the unphysical momenta are excluded by the
transformed subsidiary conditions,
\BQ
U\Phi^jU^{\dagg}=P^j,\qquad P^j\,\ketrm{phys}'=0,
\EQ
where $\ketrm{phys}'=U\ketrm{phys}$ is a transformed physical state.
Roughly speaking, this unitary transformation is interpreted
as the change of variables so that some of the new variables
parametrize gauge orbits themselves.
In other words, it separates all degrees of freedom into those of
gauge orbits and the rest. We will see this method later in more
detail by using a simple field theoretical model.
One of the advantages of the quantum Hamiltonian reduction
is that we can go to other gauges
once we choose other unitary transformations.
The transition from one gauge to another is also
implemented by a unitary transformation and the equivalence
between two gauges is shown straightforwardly.
To avoid confusion arising from the term {\it quantum reduction},
 we emphasize that the reduction of the phase space is completed in
(\ref{genecon}) and {\it quantum reduction} refers not to this
procedure but to
 extraction of a physical Hamiltonian by the unitary-transformation
method.

Although QHR has
been  applied to various gauge field theories
[\cite{GerSak}-\cite{UtiSaka},\cite{LNOT}], the effects of
regularization of composite operators have never been discussed.
The primary purpose of this paper is to examine the influence of
regularization on the reduction transformation and compare the
results with those obtained  without regularization and with those
of the classical reduction.
    The Schwinger model -- quantum electrodynamics in 1+1 dimensions with
 massless fermions -- is the simplest gauge field theory
and moreover is exactly solvable
[\cite{Sch}-\cite{Kuro}]. It has exhaustively been investigated
mainly because, besides being solvable, it is interpreted as a toy
model for QCD in four dimensional spacetime
 in that this model possesses a confining potential
and a topological structure of the vacuum (the
$\theta$-vacuum).  In the Hamiltonian formalism, we can explicitly
construct the Hilbert space and implement regularization without
resorting to any perturbative method [\cite{Manton}-\cite{Yazaki}].
Therefore the effects of a unitary transformation can be estimated
unambiguously. From these reasons, we employ the Schwinger model as
a laboratory to examine the  effects of regularization.

This paper is organized as follows: In Sec.\ II, basic ingredients of
the Schwinger model are reviewed briefly and the physical
Hamiltonian is derived by classical reduction. In Sec.\ III, the
conventional method of QHR is applied to the Schwinger model without
taking into account any regularization. In Sec.\ IV, the fermion
Fock space is constructed explicitly and regularization is
incorporated completely to show that regularization indeed
causes a modification to the Hamiltonian.
 In Sec.\ V the results obtained in Sec.\ IV are compared with those
in Sec.\ III and those of the classical reduction, and consistency
of these results is shown.  Finally in Sec.\ VI, a brief summary and
discussion are given.

\section{CLASSICAL REDUCTION OF THE SCHWINGER MODEL}

The Schwinger model is described by the  Lagrangian density
\BQ
\cL=-\frac{1}{4}F^{\mu\nu}F_{\mu\nu}+\bpsi i\Gmu D_{\mu}\psi,
\label{SL}
\EQ
where the field strength tensor is written in terms of the gauge
field $A_{\mu}$ as
$F_{\mu\nu}=\partial_{\mu}A_{\nu}-\partial_{\nu}A_{\mu}$, and the
covariant derivative is $D_{\mu}=\partial_{\mu}+ieA_{\mu}$, with $e$
being the coupling constant. The conjugate momenta of $A_1$ and
$\psi$ are $\Pi^1=F_{01}$ and $\pi_{\psi}=i\dagpsi$, respectively.
Gauss' law, defined as a constraint imposed by a Lagrange
multiplier $A_0$, is
\BQ
G(x)\equiv \partial_1\Pi_1+e\rho=0,             \label{SG1}
\EQ
with $\rho=\dagpsi\psi$ being the electric charge density.
Throughout this paper, we work in the temporal gauge, i.e., we set
$A_0=0$ and consider a fixed time, so that time dependence is
suppressed. The Hamiltonian density in this gauge is
\BQ
\cH=\half(\Pi_1)^2+\dagpsi\GF(-i\partial_1+eA_1)\psi.\label{SHam}
\EQ
We suppose that the space is compactified to a circle
with a circumference $L$
so that there exists a physical degree of freedom in the gauge field
[\cite{Manton}-\cite{IsoMura}].
We specify boundary conditions on the field variables as
$A_1(x+L)=A_1(x)$ and $\psi(x+L)= \e^{i\varphi}\psi(x)$, with
$\varphi$ being an arbitrary constant.

In the temporal gauge, there remains a local gauge symmetry.
The Hamiltonian is classically invariant under the gauge
transformations,
\BQA
\psi(x)&\rightarrow&\psi'(x)=\e^{ie\beta(x)}\psi(x),\label{Sgpcl}
\\ A_1(x)&\rightarrow&A'_1(x)=A_1(x)-
\partial_1\beta(x),\label{Sgacl}
\EQA
where $\beta$ is an arbitrary time-independent function.
Demanding that the transformed fields should also satisfy
the given boundary conditions,
we find the gauge parameter $\beta$ to be of the form
\BQ
\beta(x)=\beta_{\rm p}(x)+\beta_{\rm l}(x), \label{SGP}
\EQ
where a periodic function $\beta_{\rm p}(x)$
($\beta_{\rm p}(x+L)=\beta_{\rm p}(x)$)
gives a small gauge transformation and a linear function $\beta_{\rm l}$
with a discrete coefficient, $\beta_{\rm l}(x)=2\pi kx/eL \ (k\in\Z)$
gives a large gauge transformation.
 The large gauge transformation is characterized by an
arbitrary integer
$k$ representing the winding number of the homotopy group,
$\pi_1({\rm U}(1))=\Z$.

For an Abelian gauge theory on a circle there is precisely one
physical degree of freedom in the gauge field, namely its zero mode
\BQ
c=\frac{1}{L}\int_{0}^{L}dxA_1(x).
\EQ
Physically this is almost trivial because the gauge invariant
quantity constructed from the gauge field is  only the phase of the
Wilson loop variable, i.e., $\exp\{-ie\oint A_1(x)dx\}$. If the
space is not compact, this contribution vanishes and the gauge field
has no physical degrees of freedom.
Corresponding to the variable $c$, $\Pi^1$ has the zero mode
\BQ
p=\frac{1}{L}\int_{0}^{L}dx\Pi^1(x).
\EQ
Note that the momentum conjugate to $c$ is $Lp$.

Let us now consider the classical reduction of the Schwinger model.
We decompose the gauge field and its conjugate into the physical and
unphysical variables, i.e., the zero modes and remainders,
\BQA
A_1(x)&=&c+\tilde{A}_1(x),\\
\Pi^1(x)&=&p+\tilde{\Pi}^1(x).
\EQA
These variables are not independent
because of the Gauss' law constraint (\ref{SG1}).
Elimination of redundant variables is partly achieved
by substitution of the solution to the constraint.
It is easy to resolve Gauss' law, (\ref{SG1}),
\BQA
Q&=&\int_0^Ldx \,\rho(x)=0, \label{Qsol}\\
\tilde{\Pi}^1(x)\!&=&e\int_0^Ldy \, \rho(y)\tilde{\theta}(x-y),\label{Pisol}
\EQA
where $Q$ is the total fermionic charge and
$\tilde{\theta}(x-y)$ is the periodic step function
\BQ
\tilde{\theta}(x-y)\equiv\sum_{n\neq0}\frac{1}{2\pi i
n}\,\e^{i\frac{2\pi n}{L}(x-y)},\qquad
\frac{\partial}{\partial x}\tilde{\theta}(x-y)=\delta(x-y)-\frac{1}{L},
\EQ
with $\delta(x-y)$ being the periodic delta function
\BQ
\delta(x-y)\equiv\frac{1}{L}\sum_{n\in \Z}\e^{i\frac{2\pi n}{L}(x-y)}.
\label{pdf}
\EQ
These solutions (\ref{Qsol}) and (\ref{Pisol}) correspond,
respectively, to the zero mode
and non-zero mode parts of the Gauss' law operator.
Substituting (\ref{Pisol}) into the Hamiltonian and making a field
redefinition,
\BQ
\psi(x)\rightarrow\e^{-i\alpha(x)}\psi(x),\label{Dar}
\EQ
with
\BQ
\alpha(x)=e\int^L_0 dy\, \tilde{\theta}(x-y)A_1(y),\label{SRed}
\EQ
we obtain the reduced Hamiltonian
\BQ
\Hcl=\int dx\left\{\half p^2+\half\eta^2+
\dagpsi \GF (-i\partial_1 +ec)\psi\right\},\label{SHPcl}
\EQ
where
\BQ
\eta(x)=e\int_0^Ldy\rho(y)\tilde{\theta}(y-x).
\EQ
Eq.\ (\ref{Dar}) corresponds to the Darboux transformation in the
Faddeev-Jackiw formalism [\cite{FadJac}].
All the variables describing the Hamiltonian  $H_{\rm cl}$ are
invariant under small gauge transformations.
In this sense $H_{\rm cl}$ is described only by physical variables.

There still remains a gauge symmetry associated with  $Q$.
This gauge symmetry is global and cannot be eliminated.
 Indeed it has a physical meaning: The
electric flux always closes in the compact space and therefore the
total charge must be zero.
In addition to the global gauge symmetry, the reduced Hamiltonian (\ref{SHPcl})
 is invariant under a {\it local}  gauge  transformation,
\BQA
\psi&\rightarrow&\e^{2\pi ikx/L}\psi,\label{large1}\\
c&\rightarrow& c-\frac{2\pi k}{eL}.\label{large2}
\EQA
This symmetry under a large gauge transformation was also
present in the unreduced  Hamiltonian.
Although we have stated that $\psi$ and $c$ are physical
when they are invariant under {\it small}
gauge transformations, they are not physical variables in the strict sense
because they are subject to local gauge transformations (\ref{large1})
and (\ref{large2}).
However, we stick to this terminology for the reasons put forth in the next
section.

We have reduced the Schwinger model in the classical level.
We defer the quantization of this reduced system until Sec.\ IV
where we come back to this problem in comparison with the quantum Hamiltonian
reduction.

\section{QUANTUM REDUCTION WITHOUT REGULARIZATION}

In this section, we perform a quantum-mechanical reduction of the Schwinger
model without regularization.  As mentioned in the introduction, QHR is
based on the canonical temporal-gauge formulation.
Except for the Lagrange multiplier $A_0$, the fields are quantized
\BQ
\{\psi_\alpha(x),\dagpsi_\beta(y)\}=\delta_{\alpha
\beta}\delta_{\varphi}(x-y),\quad
[A_1(x),\Pi^1(y)]=i\delta(x-y),
\EQ
where $\alpha$ and $\beta$ are the spinor indices,
$\delta_{\varphi}(x-y)=\e^{i\frac{\varphi}{L}(x-y)}\delta(x-y)$
and $\delta(x-y)$ is the periodic delta function defined by
(\ref{pdf}).  The residual gauge transformation is
effected by the unitary operator,
\BQ
\cG[\beta]=\exp\left\{-i\int_0^Ldx
\left(-\Pi_1\frac{\partial}{\partial\tilde{x}}+e\rho\right)
\beta(\tilde{x})\right\}.
\EQ
The space index with a tilde, $\tilde{x}$, is defined by
\BQ
x=\left[\frac{x}{L}\right]L+\tilde{x},
\EQ
where the square brackets stand for the Gauss notation, i.e.,
$[x/L]$ is the integer part of $x/L$. By definition $0\leq
\tilde{x} <L$.  The use of $\tilde{x}$ in place of $x$ in the
integrand is required to avoid undesired contributions from the
boundaries and to reproduce the gauge transformations (\ref{Sgpcl})
and (\ref{Sgacl}). Note that for $\beta(x)$ given by (\ref{SGP}),
$\e^{ie\beta(\tilde{x})}=\e^{ie\beta(x)}$.  For periodic field operators,
e.g., $\Pi_1(\tilde{x})=\Pi_1(x)$ and $\rho(\tilde{x})=\rho(x)$.
With the decomposition (\ref{SGP}),
$\cG[\beta]=\cG[\beta_{\rm p}]\cG[\beta_{\rm l}]$.
In
\BQ
\cG[\beta_{\rm p}]=\exp\left\{-i\int_0^Ldx
G(x)\beta_{\rm p}(x)\right\},
\EQ
the Gauss' law  operator is the generator of small gauge transformations.
On the other hand,
\BQ
\cG[\beta_{\rm l}]=\exp\left\{-\frac{2\pi
ik}{eL}\int_0^Ldx(-\Pi_1(x)+e\tilde{x}\rho(x) )\right\}\label{Sgau}
\EQ
generates large gauge transformations.  The gauge invariance of the
Hamiltonian  is equivalent with the invariance under the unitary
transformation $\cG[\beta]$,
\BQ
\cG[\beta]\cH\cG^{\dagg}[\beta]=\cH.\label{SHinv}
\EQ
To eliminate this gauge symmetry from the total Hilbert space
constructed with superfluous degrees of freedom,
we impose a subsidiary condition,
\BQ
G(x)\ketrm{phys}=0.\label{subcon}
\EQ
In the canonical temporal-gauge formulation,
the redundancies associated with the {\it large} gauge transformations
cannot be eliminated because Gauss' law (\ref{subcon})
removes only the small gauge symmetries.
Therefore within the framework of the canonical temporal-gauge formulation,
{\it physical} variables are variables invariant under the {\it small}
gauge transformations.  In this sense, we call $\psi$ and $c$ physical
and $\tilde{A}_1$ unphysical.

Physical predictions can be made by the constrained
Hamiltonian (\ref{SHam}) under the subsidiary condition
(\ref{subcon}), but by exploiting an appropriate unitary transformation
 we can go to a more transparent representation,
where the subsidiary condition becomes trivial and
the Hamiltonian is decomposed definitely
into physical and unphysical parts.
Consider a unitary operator $U$ defined by
\BQ
U=\e^{i\int dx \rho(x)\alpha(x)},\label{SU}
\EQ
where $\alpha(x)$ was given by (\ref{SRed}).  The fundamental field
operators are transformed as
\BQA
U\psi(x)U^{\dagg}&=&\e^{-i\alpha(x)}\psi(x),\label{SUpsi}\\
UA_1(x)U^{\dagg}&=&A_1(x),\\
U\Pi_1(x)U^{\dagg}&=&\Pi_1(x)+\eta(x).\label{SUPi}
\EQA
To prove (\ref{SUpsi}) we have used the fact that $\alpha(x)$ is a
periodic function. From (\ref{SUpsi})-(\ref{SUPi}), we obtain the transformed
Hamiltonian,
\BQA
U\cH U^{\dagg}
      &=&\half(\Pi_1+\eta)^2+\dagpsi\GF(-i\partial_1+ec)\psi \NN
      &=&\cH_{\rm C}+\cH'_{\rm C},         \label{SHnai1}
\EQA
with
\BQA
\cH_{\rm C}&\equiv& \half p^2 + \half \eta^2
	    + \dagpsi\GF(-i\partial_1+ec)\psi,   \label{SHP} \\
\cH'_{\rm C}&\equiv& \half  \tilde{\Pi}_1 (\tilde{\Pi}_1 +
2\eta),\label{SHunP}
\EQA
where the suffix C means that the transformed quantities
belong to the Coulomb-gauge representation.
The Gauss' law operator becomes
\BQA
G_{\rm C}(x) \equiv UG(x)U^{\dagg}
			      &=&\partial_1(\Pi_1+\eta)+e\rho\NN
			      &=&\partial_1 \Pi_1 +e\frac{Q}{L}.
\label{SG2}
\EQA
This picks up the physical states,
\BQ
\left(\partial_1\Pi_1+e\frac{Q}{L}\right)\ketrm{phys}'=0\ \ \
\Longrightarrow\ \ \
\left\{\begin{array}{l}
		Q\ketrm{phys}'=0\\
		\tilde{\Pi}_1\ketrm{phys}'=0
       \end{array}  \right. \ ,\label{trsubcon1}
\EQ
where $\ketrm{phys}'=U\ketrm{phys}$ is the transformed physical state.
The charge $Q$ was present in the classical reduction
as the generator of the global gauge transformation.

The transformed unitary operator for gauge transformations is
\BQ
\cG_{\rm C}[\beta]\equiv U\cG\,[\beta]\,U^{\dagg}
=\e^{-i\int_0^L dyG_{\rm C}(x)\beta_{\rm p}(y)}
\cG[\beta_{\rm l}],
\label{SgauR}
\EQ
where $\cG[\beta_{\rm l}]$, defined by (\ref{Sgau}), is not modified
by the unitary transformation owing to the identity $\int_0^L\eta(x)dx=0$.
$\cG_{\rm C}[\beta]$ brings about the transformations,
\BQA
\lefteqn{\cG_{\rm C}[\beta]\psi(x)\cG^{\dagg}_{\rm C}[\beta]
=\e^{i\frac{e}{L}\int_0^L \beta_{\rm p}(y)dy+2\pi ikx/L}\psi(x),}
\label{SgauRpsi}\\
\lefteqn{\cG_{\rm C}[\beta]\,c\,\cG^{\dagg}_{\rm C}[\beta]
=c-\frac{2\pi k}{eL},} \\
\lefteqn{\cG_{\rm C}[\beta]\tilde{A}_1(x)\cG^{\dagg}_{\rm C}[\beta]
=\tilde{A}_1(x)-\partial_1\beta_{\rm p}(x),} \label{SgauRa}
\EQA
whereas $p$ and $\tilde{\Pi}^1$ are invariant.
Note that for small gauge transformations ($\cG[\beta_{\rm l}]=1$),
only $\tilde{A}_1$ changes locally.

Furthermore,  we obtain in the physical subspace
\BQ
\left(\cH_{\rm C}+\cH'_{\rm C}\right)\ketrm{phys}'
		      =\cH_{\rm C}\ketrm{phys}'.
\EQ
When limited to the physical states, this result is the same as that
of the classical reduction (\ref{SHPcl}).  The transformed
Hamiltonian (\ref{SHnai1}) is invariant under gauge
transformations generated by the {\it transformed}  Gauss' law
operator (\ref{SG2}).

We have been able to extract the physical Hamiltonian $\cH_{\rm
C}$ from the constrained one (\ref{SHam}) by the unitary
transformation (\ref{SU}). The key of QHR is that  the following
two things occur simultaneously; elimination of  redundant variables
from a Hamiltonian and simplification of  constraints.  While the
redundant {\it coordinates} vanish from the Hamiltonian,  the {\it
momenta} conjugate to them are excluded  by the transformed
constraints. Hence all the redundant variables are removed from the
Hamiltonian.  In the classical reduction in Sec.\ II, the
above two things were performed step by step, resolution of
Gauss' law, (\ref{Qsol})-(\ref{Pisol})  and redefinition of the fermion
field (\ref{Dar}).

Before leaving this section it would be interesting to comment on
another meaning of the unitary transformation $U$ [\cite{ChLee}].
In the Schr\"odinger representation in which $A_1$ is diagonal, i.e.,
$\Pi_1(x)=i\delta/\delta A_1(x)$, (\ref{subcon}) is rewritten as
a functional-derivative equation,
\BQ
\left( i\partial_1\frac{\delta}{\delta
\tilde{A}_1}+e\rho\right)\Psi[A_1]=0,\label{fde}
\EQ
where $\Psi[A_1]=\bra{A_1}{\rm phys}\rangle$ is a wavefunctional of
a state $\ketrm{phys}$. Eq.\ (\ref{fde}) is solved for
\BQ
\Psi[c,\tilde{A}_1]=U[\tilde{A}_1]\Phi[c],
\EQ
where $U$ is identical with the unitary operator  (\ref{SU})
and $\Phi[c]$ depends only on $c$.
Note that the unitary operator $U$ is independent of $c$ because
$\alpha(x)$ defined in (\ref{SRed}) does not contain $c$
(remember that $\tilde{\theta}(x)$ is the {\it
periodic} step function). The condition that $\Phi[c]$ should
include only $c$ can be expressed as
\BQ
\frac{\delta}{\delta \tilde{A}_1}\Phi[c]=0.
\EQ
This is nothing but the transformed subsidiary condition
$\tilde{\Pi}_1\ketrm{phys}'=0$
with the identification $\Phi[c]=\bra{A_1}{\rm phys}\rangle'$.
Therefore the unitary transformation $U$ is also interpreted
as the wavefunctional for the unphysical gauge field.


\section{QUANTUM REDUCTION WITH REGULARIZATION}

When we deal with a quantum field theory, we must regularize
composite operators. The reflection of regularization can be
observed, for example, in the current algebra as an anomalous
term called the `Schwinger term' [\cite{Jac2}].  This extra term must
have influence on the reduction transformations.  Consider a
transformation of the fermion current
$J_{\mu}=\dagpsi\gmu\psi$.
If we perform the unitary transformation $U$ defined by (\ref{SU})
 {\it without} regularization, we have
\BQA
UJ_{\mu}U^{\dagg}&=&U\dagpsi U^{\dagg}\gmu U\psi U^{\dagg}\NN
&=&J_{\mu}.\nonumber
\EQA
On the other hand, for regularized currents, we have
\[
UJ_{\mu}^\reg(x)U^{\dagg}=J_{\mu}^\reg(x)
  +i\int dy[J_0^\reg(y),J_{\mu}^\reg(x)]\alpha(y)+\cdots.
\]
Therefore $UJ_0^\reg U^{\dagg}=J_0^\reg$ but
$UJ_1^\reg U^{\dagg}\neq J_1^\reg$ because of the Schwinger term.
Since the Hamiltonian density (\ref{SHam})
contains the fermion current in the interaction term with the gauge
field ($-eJ_1A_1$), it also transforms non-trivially at least for
the interaction term.  Therefore QHR should be carefully performed
with regularization of composite operators such as fermion
currents, charge and Hamiltonian.

\subsection{Construction of the fermion Fock space and
regularization}

We construct the fermion Fock space to evaluate regularization explicitly.
We essentially follows Iso and Murayama [\cite{IsoMura}].
The fermionic part of the Hamiltonian density (\ref{SHam}) can be expressed as
\BQ
\cH_{\rm F}=\dagpsi
   \left(\matrix{h_{\rm F}&0\cr 0 & -h_{\rm F}\cr} \right)
   \psi,
\EQ
where $h_{\rm F}=-i\partial_1+eA_1$ and we have used
the chiral representation for the $\gamma$-matrices:
$\gamma^0=\sigma_1$, $\gamma^1=-i\sigma_2$
and $\GF=\gamma^0\gamma^1=\sigma_3$, where $\sigma_i$ is the Pauli matrix.
Solving the eigen-equations
$h_{\rm F}\chi_n(x)=\varepsilon_n\chi_n(x)$, we obtain
\BQ
\chi_n(x)=\frac{1}{\sqrt{L}}\e^{-ie\int_0^x A_1(y)dy+i\varepsilon_n
x},\qquad
\varepsilon_n=\frac{2\pi}{L}\left(n+
\frac{\varphi}{2\pi}+\frac{ecL}{2\pi}\right)
\EQ
for $n\in\Z$.  Note that the energy spectrum depends only on $c$.

Next we expand $\psi$ in the orthonormal complete set $\{\chi_n\}$,
\BQ
\psi(x)=\sum_{n\in\Z}\left\{a_n\chi_n(x)\left(\matrix{1\cr0
\cr}\right)
		  + b_n\chi_n(x) \left(\matrix{0 \cr 1\cr}\right) \right\},
\EQ
and impose anti-commutation relations,
\BQ
\{ a_n,a_m^{\dagg}\}=\delta_{nm},\ \ \ \{b_n,b_m^{\dagg}\}=\delta_{nm}.
\EQ
The fermion Hamiltonian $H_{\rm F}=\int_0^L\cH_{\rm F}dx$ is diagonalized
in terms of these operators and decomposed into the positive- and
negative-chirality pieces as
\BQ
H_{\rm F}=H_++H_-,
\EQ
\BQ
H_+=\sum_{n\in\Z}\varepsilon_na_n^{\dagg}a_n,\qquad
H_-=\sum_{n\in\Z}(-\varepsilon_n)b_n^{\dagg}b_n.
\EQ
We define $N_\pm$-vacua by
\BQA
\lefteqn{\ket{{\rm
vac};N_+}_+\equiv\prod_{n=-\infty}^{N_+-1}a_n^{\dagg}
\ket{0}\ ,\qquad N_+\in\Z,}\\
\lefteqn{\ket{{\rm vac};N_-}_-\equiv
\prod^{\infty}_{n=N_-}b_n^{\dagg}\ket{0}\ ,\qquad N_-\in\Z,}
\EQA
where $\ket{0}$ is the state of nothing defined by
$a_n\ket{0}=b_n\ket{0}=0$ for
$\forall n\in\Z$.
Since the positive- and negative-chirality sectors are decoupled,
 the total fermion vacua are given by
\BQ
\ket{{\rm vac};N_+,N_-}=\ket{{\rm vac};N_+}_+\ket{{\rm vac};N_-}_-.
\EQ
For some fixed $N_{\pm}$, we obtain the fermion Fock space by acting
creation or annihilation operators on the $N_{\pm}$-vacua.

Since the composite operators such as $H_\pm$  become
divergent on the $N_\pm$-vacua, we regularize them by the
$\zeta$-function regularization,
\BQA
\lefteqn{H^\reg_+=\lim_{s\rightarrow0}\sum_{n\in\Z}\varepsilon_n
a_n^{\dagg}a_n\frac{1}{(\lambda \varepsilon_n)^s},}\\
\lefteqn{H^\reg_-=\lim_{s\rightarrow0}\sum_{n\in\Z}(-\varepsilon_n)
b_n^{\dagg}b_n\frac{1}{(\lambda \varepsilon_n)^s},}
\EQA
where $\lambda$ is an arbitrary constant with the dimension of length.
It is easy to check that {\it this regularization respects gauge symmetry}.
We also regularize the current operators
$J_{\pm}(x)\equiv\half\left(J_0(x)\pm J_1(x)\right)$ by
\BQA
j_+^n=\sum_{q\in \Z}a_q^{\dagg}a_{q+n}
&\rightarrow
& ( j_+^n)^\reg
      =\lim_{s\rightarrow 0} \sum_{q\in \Z}
    a_q^{\dagg}a_{q+n}\frac{1}{(\lambda\varepsilon_q)^s},
    \label{regcurrentplus}\\
j_-^n=\sum_{q\in \Z}b_{q+n}^{\dagg}b_q
	    &\rightarrow
	    & (j_-^n)^\reg
		       =\lim_{s\rightarrow 0} \sum_{q\in \Z}
				     b_{q+n}^{\dagg}b_q \frac{1}{(\lambda\varepsilon_q)^s},
									   \label{regcurrentminus}
\EQA
where $j_\pm^n$ are the Fourier transforms defined by
$J_\pm(x)=L^{-1}\sum_{n\in\Z}j_\pm^n\exp(\mp i\frac{2\pi n}{L}x)$.
Here we have adopted the $\zeta$-function method, but we can resort
to any other regularizations, e.g., the heat-kernel or the
point-splitting regularization [\cite{Manton},\cite{Jac2}], as long
as they respect the gauge symmetry. Using the formulae for
the $\zeta$-function,
$
\zeta(s,a)=\sum_{n=0}^{\infty}(n+a)^{-s},
$
\BQ
\zeta(0,a)=\half-a,\ \ \ \ \ \zeta(-1,a)
		=-\half\left(\half - a\right)^2+\frac{1}{24},
\EQ
we can evaluate the eigen-values of the regularized operators.
For example, we obtain
\BQ
Q_{\pm}^\reg\ket{{\rm vac};N_{\pm}}_{\pm}
     =\pm \left( N_{\pm}+ \frac{ecL}{2\pi} \right)\ket{{\rm
vac};N_{\pm}}_{\pm},\label{SQvev}
\EQ
\BQ
H_{\pm}^\reg\ket{{\rm vac};N_{\pm}}_{\pm}
 =\frac{2\pi}{L}\left(\half \langle Q_{\pm}^\reg\rangle^2-\frac{1}{24}
\right)\ket{{\rm vac};N_{\pm}}_{\pm}, \label{SHvev}
\EQ
where $Q_{\pm}^\reg= (j^{n=0}_{\pm})^\reg$ are the fermionic charge operators.
As previously mentioned, the current algebra is modified by the
regularization (see Appendix),
\BQ
[(j_{\pm}^n)^\reg,(j_{\pm}^m)^{\reg\dagg}]=n\delta_{nm},\label{Sterm1}
\EQ
or in the coordinate space,
\BQ
[J_{\pm}^\reg(x), J_{\pm}^\reg(y)]=\pm
\frac{i}{2\pi}\frac{\partial}{\partial x}\delta(x-y).\label{jjpm}
\EQ
{}From (\ref{jjpm}), we have
\BQA
[J_0^\reg(x),J_1^\reg(y)]
&=& \frac{i}{\pi}\frac{\partial}{\partial x}\delta(x-y),\label{Sterm2}\\
{}[J_0^\reg(x), J_0^\reg(y)]&=&[J_1^\reg(x), J_1^\reg(y)]=0. \label{others}
\EQA
Note that only the first relation (\ref{Sterm2})
has the extra term originating from regularization.

By commuting $H_{\pm}^\reg$ with $(j_{\pm}^n)^{\reg\dagg}$,
we have
\BQ
[H_{\pm}^\reg,    (j_{\pm}^n)^{\reg\dagg}]
	       =\frac{2\pi n}{L}(j_{\pm}^n)^{\reg\dagg},\label{SHj}
\EQ
and this leads to
\BQ
[H_{\pm}^\reg,J_{\pm}^\reg(x)]=\mp i\partial_1J_{\pm}^\reg(x).
\label{SHJ}
\EQ
This relation and the current algebra (\ref{jjpm}) are the
indispensable tools for the calculation of the reduction
transformation.

\subsection{Reduction transformation}

Now we are ready to make a unitary transformation of regularized
operators. As a unitary operator, we
employ $U$ defined in (\ref{SU}). It is divided into positive- and
negative-chirality parts
\BQA
U&=&\e^{i\int dx J_+(x)\alpha(x)}\e^{i\int dx J_-(x)\alpha(x)}\NN
	 &\equiv&U_+U_-.
\EQA
For the positive-chirality sector, its fermion Hamiltonian is transformed as
\BQA
U_+H_+^\reg U_+^{\dagg}
&=&H_+^\reg+ i\int dx [J_+^\reg(x),H_+^\reg]\alpha(x)\NN
&   & +  \frac{1}{2!} i^2 \int \int dxdy
		      [J_+^\reg(y),[J_+^\reg(x),H_+^\reg]]
				  \alpha(x)\alpha(y)+\cdots    \NN
&=&H_+^\reg+ \int dx J_+^\reg(x)\partial_1\alpha(x)
	 +\frac1{4\pi}\int
dx\partial_1\alpha(x)\partial_1\alpha(x).\label{UHUp}
\EQA
where we have used the relations (\ref{SHJ}) and (\ref{Sterm1}).
All composite operators must be regularized but regularization of the
gauge fields does not affect our results.  Note that, because of the
Schwinger term, the third term turns out
not to vanish in contrast to the reduction without regularization.
 Likewise, the
negative-chirality fermion Hamiltonian is transformed as
\BQ
U_-H_-^\reg U_-^{\dagg}
     =H_-^\reg- \int dx J_-^\reg(x)\partial_1\alpha(x)
+\frac1{4\pi}\int
dx\partial_1\alpha(x)\partial_1\alpha(x).\label{UHUm}
\EQ
Combining (\ref{UHUp}) and (\ref{UHUm}), we obtain for
the total fermion Hamiltonian,
\BQA
UH_{\rm F}^\reg U^{\dagg}
	&=&H_{\rm F}^\reg+\int dx\, \left(J_1^\reg\partial_1\alpha
	 +\frac1{2\pi}\left(\partial_1\alpha\right)^2\right)\NN
	&=&H_{\rm F}^\reg+\int dx\,\left(eJ_1^\reg\tilde{A}_1
	 +\half m^2\tilde{A}_1^2\right),
\EQA
where $m=e/\sqrt{\pi}$ and we have used $\partial_1\alpha=e\tilde{A}_1$.
To get a transformation law for the total Hamiltonian,
we have to know a commutator between the kinetic term of the gauge field,
$\half\Pi_1^2$, and the regularized fermion charge density $J^\reg_0$,
but the commutator does not change by regularization.
(Although Gauss' law relates $\Pi_1$ with $\rho=J_0$, the commutator
between $J_0$'s does not get modified. See Eq.\ (\ref{others}).)
Eventually the result of the unitary transformation
 for the total Hamiltonian is
\BQ
H_{\rm C}\equiv UH^\reg U^{\dagg}
	=  H_{\rm F}^\reg+ \int dx\, \left\{\half \left(\Pi_1 + \eta \right)^2
	 + eJ_1^\reg\tilde{A}_1
	 +  \half m^2 \tilde{A}_1^2\right\}.             \label{SHreg}
\EQ
We have introduced the unitary transformation $U$ to eliminate
non-zero modes $\tilde{A}_1$ from the Hamiltonian, as is seen in
Eq.\ (\ref{SHnai1}), but it has turned out that with regularization
$H_{\rm C}$ contains $\tilde{A}_1$.

Since the Gauss' law operator is not modified by regularization, the physical
state is specified by the same conditions as those of the classical
reduction:
\BQ
\tilde{\Pi}_1 \ketrm{phys}'=0,\ \ \ \ Q\ketrm{phys}'=0.\label{trsubcon}
\EQ
 As stated in Sec.\ III, in the Schr\"odinger representation
where $A_1$ is diagonal, (\ref{trsubcon}) requires that the
physical states should not depend on $\tilde{A}_1$. If we are
allowed to write the Hamiltonian $H_{\rm C}$ as
\BQ
H_{\rm C}=\int dx\left\{\half\left(\Pi_1+\eta\right)^2+
\dagpsi\GF(-i\partial_1+ec)\psi+\half m^2\tilde{A}_1^2\right\}^\reg,
\EQ
it is evident that only the last term is an extra additional to the
result (\ref{SHnai1}).

\subsection{Hamiltonian in terms of the currents}

For later convenience we rewrite the Hamiltonian
in terms of the fermion currents.
To this end, we notice that operators
$B_{\pm}^n\equiv\frac{1}{\sqrt{n}}j_{\pm}^n \ (n>0)$ satisfy the
{\it bosonic} commutator
\BQ
[B_{\pm}^n,B_{\pm}^{m\dagg}]=\delta_{nm},\label{bosonic}
\EQ
and from Eq.\ (\ref{SHj}) we have
\BQ
[H_{\rm F},B_{\pm}^{n\dagg}]=\frac{2\pi n}{L}B_{\pm}^{n\dagg}.
\EQ
These relations and the vacuum expectation value of $H_{\rm F}$,
 (\ref{SHvev}),
indicate that $H_{\rm F}$ can be expressed
 in terms of the currents as [\cite{IsoMura}]
\BQ
H_{\rm F}=\frac{2\pi}{L}
  \left\{\half (Q_+^2+Q_-^2) - \frac{1}{12} +
 \sum_{n>0}\left(j_+^{n\dagg}j_+^n + j_-^{n\dagg}j_-^n \right)
\right\}.\label{Hfcurrent}
\EQ
It is a direct consequence of regularization that we can express
the fermion Hamiltonian in terms of the currents.  Thus when we write
$H_{\rm F}$ in terms of the fermion currents without the suffix `reg,'
we take it as a matter of course that the currents are regularized.
By acting the creation operators
$B_{\pm}^{n\dagg}$ on the $N_{\pm}$-vacua,
we obtain the Fock space that is equivalent to the one constructed
by $a^{(\dagg)}_n$ and $b^{(\dagg)}_n$. Observing that
\BQ
\half\int dx\eta^2(x)=\frac{1}{2L}\sum_{n\neq
0}\left(\frac{eL}{2\pi
n}\right)^2\left(j_+^n+j_-^{n\dagg}\right)\left(j_+^{n\dagg}+j_-^n\right),
\EQ
we can rewrite the total Hamiltonian
\BQA
H_{\rm C}&=&UHU^{\dagg}=h_{\rm C}+h'_{\rm C},\label{SHredT}\\
h_{\rm C}&=&-\frac{1}{2L}\frac{\partial^2}{\partial c^2}+
\frac{2\pi}{L}\left\{\frac{1}{2}(Q_+^2+Q_-^2)-\frac{1}{12}\right\}\NN
	& &+\frac{1}{2L}\sum_{n\neq 0}
\left(\frac{eL}{2\pi n}\right)^2\left(j_+^n+j_-^{n\dagg}\right)
\left(j_+^{n\dagg}+j_-^n\right)\NN
	& &+\frac{2\pi}{L}\sum_{n>0}\left(j_+^{n\dagg}j_+^n
	       +j_-^{n\dagg}j_-^n\right),\label{SIM}
\EQA
\BQ
h'_{\rm C}=\int_0^Ldx\left\{\half
\tilde{\Pi}_1(\tilde{\Pi}_1+2\eta)+eJ_1\tilde{A}_1+\half m^2
\tilde{A}_1^2  \right\},
\EQ
where  $Lp=-i\partial/\partial c$ was used.
The first part of the transformed Hamiltonian, $h_{\rm C}$, has
exactly the same form as that of  Ref.\ [\cite{IsoMura}], in which
Gauss' law is solved explicitly after quantization.

On the other hand, $h'_{\rm C}$ is estimated on the physical state,
\BQ
h'_{\rm C}\ketrm{phys}'
  =\int dx \left( eJ_1\tilde{A}_1 + \half m^2
\tilde{A}_1^2\right)\ketrm{phys}'.
\EQ
Although the physical state does not depend on  $\tilde{A}_1$,
 we cannot ignore $h'_{\rm C}$:  If limited to the physical
state,
$\tilde{A}_1$ seems to serve as an auxiliary field
generating current-current interaction.
 Therefore one might be tempted to conclude that our results disagree
with those of Ref.\ [\cite{IsoMura}].  However we should not compare them
at this stage. In the next section one sees that our results turn out to be
consistent.

The comparison should be made with the regularized Hamiltonian
that was reduced classically.  We regularize the classically
reduced Hamiltonian $\Hcl$ (see (\ref{SHPcl})) following the
same way as that of the unreduced case.  The important point to be
noted is that unlike the preceding case, the first quantization is
performed for the fermion Hamiltonian of $\Hcl$, i.e.,
$(\Hcl)_{\rm F}=\int dx\dagpsi\GF(-i\partial_1+ec)\psi$,
where only the zero mode of the gauge field is coupled to the
fermion field.  In this case, the regularization respects a {\it
residual} gauge symmetry.  Representing $\Hcl$ with the regularized currents,
we obtain a Hamiltonian of the same form as that of
Ref.\ [\cite{IsoMura}]. For clarity, we give its expression again at
the risk of being tedious,
\BQA
(\Hcl)^\reg
   &=&-\frac{1}{2L}\frac{\partial^2}{\partial c^2}+\frac{2\pi}{L}
\left\{\frac{1}{2}(Q_+^2+Q_-^2)-\frac{1}{12}\right\}\NN
   & &+\frac{1}{2L}\sum_{n\neq 0}\left(\frac{eL}{2\pi n}\right)^2
\left(l_+^n+l_-^{n\dagg}\right)\left(l_+^{n\dagg}+l_-^n\right)\NN
	& &+\frac{2\pi}{L}\sum_{n>0}\left(l_+^{n\dagg}l_+^n
     +l_-^{n\dagg}l_-^n\right),\label{SIMclas}
\EQA
where we denote the currents as $l_\pm^n$ to distinguish them  from $j_\pm^n$.

{}From Eq.\ (\ref{SQvev}),
the zero mode part of the Hamiltonian is estimated on the physical state as
\BQA
H^{\rm zero}&=&-\frac{1}{2L}\frac{\partial^2}{\partial c^2}+\frac{2\pi}{L}
\frac{1}{2}(Q_+^2+Q_-^2)\NN
&=&-\frac{1}{2L}\frac{\partial^2}{\partial c^2}+\half m^2L\left(c+\frac{2\pi
N}{eL}\right)^2,\label{zeroH}
\EQA
where $N=N_+=N_-$ from the requirement that the total charge must be zero,
$Q=N_+-N_-=0$. Note that this Hamiltonian contains the mass term for $c$.


\section{CONSISTENCY OF THE CLASSICAL AND QUANTUM REDUCTIONS}

The unitary transformation $U$ has yielded a new Hamiltonian $H_{\rm C}$,
(\ref{SHreg}), and a new Gauss' law operator $G_{\rm C}$, (\ref{SG2}).
It is well known that the regularization produces a restoring force for the
zero mode $c$, but we have now discovered that the
regularization has also   produced a mass term for the non-zero mode
$\tilde{A}_1$.  These results  apparently contradict with those
without regularization.  In this section we pin down the origin of
this apparent discrepancy and prove somewhat heuristically that
these two approaches are nonetheless consistent.

To begin with, let us observe the strange behavior of $H_{\rm C}$
under gauge transformations. Its accurate understanding gives a
key to the problem of the equivalence between quantum and classical
reductions.

The gauge transformations of the fundamental operators was given in
(\ref{SgauRpsi})-(\ref{SgauRa}).
It follows immediately that our result (\ref{SHreg}) is not gauge invariant
classically because the mass term is present for $\tilde{A}_1$.
Nevertheless the invariance of the {\it unreduced} Hamiltonian
($\cG H\cG^{\dagg}=H$)
 ensures that the new Hamiltonian (\ref{SHreg}) is also invariant:
\BQ
\cG_{\rm C}[\beta]\, H_{\rm C}\, \cG^{\dagg}_{\rm C}[\beta]=H_{\rm
C}.\label{SHinvR}
\EQ
Indeed this can be shown explicitly as a {\it regularized} equation
if we notice that
\BQ
[\Pi_1(x), J_1^\reg(y)]=-i\frac{e}{\pi}\left(\delta(x-y)-
\frac{1}{L}\right).\label{Spij}
\EQ
In Appendix we derive this within the framework of the $\zeta$-function
regularization.

On the other hand, the result of the unitary transformation
without regularization, (\ref{SHnai1})-(\ref{SHunP}),
is gauge invariant classically,
but is not invariant if we consider regularization.
What caused  the violation of the classical gauge symmetry?
The reason is self-evident.  It is true that our regularization respects
the gauge symmetry generated by $G$, but after the unitary transformation,
the Gauss' law operator changes into $G_{\rm C}$
and therefore the gauge transformation is different from the original one.
On the contrary, the reduction was performed by the unitary transformation
 under the regularization
that respects the original gauge symmetry $\cG$.
 Thus there is no reason why the transformed Hamiltonian possesses
the classical gauge symmetry.

This can be viewed more transparently in the point-splitting regularization.
Since our regularization preserves the original gauge symmetry,
 the currents are regularized as
\BQ
J_{\mu}^\reg(x)=\lim_{\varepsilon\rightarrow
0}\left\{\bpsi(x+\varepsilon)\gmu\exp\left(ie\int_{x}^{x+\varepsilon}dy\,A_1(y)\right)\psi(x)- {\rm v.e.v.}\right\},
\EQ
where the line integral is inserted to preserve
the gauge symmetry under $\cG$.
On the other hand, if we want to preserve the gauge symmetry
$\cG_{\rm C}$,  we should regularize the currents as
\BQ
\cJ_\mu^\reg(x)\equiv\lim_{\varepsilon\rightarrow0}
\left\{\bpsi(x+\varepsilon)\gmu\exp
\left(ie\int_{x}^{x+\varepsilon}dy\,c\right)
\psi(x)-{\rm v.e.v.}\right\}.\label{Spoint2}
\EQ
Classically  these currents are invariant under either of the
 gauge transformations $\cG$ and $\cG_{\rm C}$.
 If we consider the regularization, however, the current $J_1^\reg$
 is transformed as
\BQA
\cG[\beta]J_1^\reg\cG^{\dagg}[\beta]&=& J_1^\reg, \label{SJgau1}\\
\cG_{\rm C}[\beta]J_1^\reg\cG^{\dagg}_{\rm C}[\beta]&=&
J_1^\reg+\frac{e}{\pi}\partial_1\beta_{\rm p}.\label{SJgau2}
\EQA
By contrast, $\cJ_1^\reg$ behaves as
\BQA
\cG[\beta]\cJ_1^\reg\cG^{\dagg}[\beta] &=&
 \cJ_1^\reg-\frac{e}{\pi}\partial_1\beta_{\rm p},\\
\cG_{\rm C}[\beta]\cJ_1^\reg\cG^{\dagg}_{\rm C}[\beta] &=& \cJ_1^\reg.
\EQA
Therefore $J_1^\reg$ and $\cJ_1^\reg$ are invariant under $\cG$ and
$\cG_{\rm C}$, respectively. For definiteness, let us denote the
regularization that respects the original gauge symmetry as $\RG$,
and the one that respects the gauge symmetry generated by $\cG_{\rm
C}$ as $\RGC$.

The regularization $\RGC$ respects the residual global symmetry of
the classically reduced Hamiltonian,
\BQ
\cG_{\rm C}[\beta](\Hcl)^\reg
\cG^{\dagg}_{\rm C}[\beta]=(\Hcl)^\reg.
\EQ
Therefore the Fourier components of $\cJ_{\pm}^\reg$ in the
point-splitting regularization, (\ref{Spoint2}), should be
identified with $l^n_{\pm}$ that appeared in $(\Hcl)^\reg$,
(\ref{SIMclas}),
\BQ
\cJ_{\pm}^\reg(x)=\frac{1}{L}
\sum_{n\in\Z}(l^n_\pm)^\reg\e^{\mp i\frac{2\pi n}{L}x}.
\EQ
Here one must be aware that the conceptually different
two currents $(l^n_{\pm})^\reg$ and $(j_\pm^n)^\reg$ are identical in the
$\zeta$-function regularization since the fermion energy
$\varepsilon$ depends only on the zero mode of the gauge field.
The difference was absorbed into the basis function $\chi_n$.
Now we can understand the coincidence of the Hamiltonian in Ref.\
[\cite{IsoMura}] with our result (\ref{SIMclas}). It is accidental
that $H_{\rm F}$ and $(\Hcl)_{\rm F}$ are represented in the
same form. Indeed $H_{\rm F}$ does contain the unphysical field and
$(\Hcl)_{\rm F}$ does not.
Therefore the result of Ref.\,[\cite{IsoMura}] is insufficient in
that all the unphysical degrees of freedom are
not eliminated from the Hamiltonian.

We have learned that currents in different regularization schemes
 cannot be identified straightforwardly.
 Therefore it is not meaningful
to discuss the equivalence of the Hamiltonians by comparing only their forms.
 We must find an appropriate current
to be compared with that of the classically reduced case.

{}From (\ref{SJgau2}), we see that
the current $J_1^\reg$ (regularized by $\RG$) is not invariant
under the residual gauge transformation.
By now, we have constructed this system
 in terms of  gauge {\it variant} variables unintentionally.
 We need to reformulate it in terms of  gauge {\it invariant} variables.
Transforming Eq.\ (\ref{SJgau1}) by $U$,
 we obtain $\cG_{\rm C}(U J^\reg_1 U^{\dagg}) \cG^{\dagg}_{\rm C}=
	U J^\reg_1 U^{\dagg} . $
Thus $U J^\reg_1 U^{\dagg}=J_1^\reg+\frac{e}{\pi}\tilde{A}_1$ is gauge
invariant under $\cG_{\rm C}$
and we can define a gauge-invariant current by
\BQ
\hat{J}^\reg_1\equiv J_1^\reg+\frac{e}{\pi}\tilde{A}_1. \label{SJinv}
\EQ
Since the physical state does not contain $\tilde{A}_1$, we are
allowed to identify these two currents $\hat{J}^\reg_1$ and
$J_1^\reg$ in the physical subspace.  Let us rewrite our result
(\ref{SHredT}) in terms of the gauge-invariant current. The Fourier
components of
$\hat{J}^\reg_{\pm}=J^\reg_{\pm}\pm \frac{e}{2\pi}\tilde{A}_1$
is written as
\BQ
(\hat{j}_\pm^n)^\reg
=(j_\pm^n)^\reg+\frac{e}{2\pi}\tilde{A}^{\mp n},\label{jhatn}
\EQ
where $\tilde{A}^n$ is the Fourier component of $\tilde{A}_1(x)$,
\[
\tilde{A}_1(x) =\frac{1}{L}\sum_{n\neq 0}\tilde{A}^n \e^{i\frac{2\pi n}{L}x}.
\]
Inserting (\ref{jhatn}) into (\ref{SHredT}), we finally obtain the
desired result:
\BQA
H_{\rm C}&=&-\frac{1}{2L}\frac{\partial^2}{\partial c^2}+
\frac{2\pi}{L}\left\{\frac{1}{2}(Q_+^2+Q_-^2)-\frac{1}{12}\right\}\NN
& &+\frac{1}{2L}\sum_{n\neq 0}\left(\frac{eL}{2\pi n}\right)^2
\left(\hat{j}_+^n+\hat{j}_-^{n\dagg}\right)
\left(\hat{j}_+^{n\dagg}+\hat{j}_-^n\right)\NN
& &+\frac{2\pi}{L}\sum_{n>0}\left(\hat{j}_+^{n\dagg}\hat{j}_+^n
      +\hat{j}_-^{n\dagg}\hat{j}_-^n\right)\NN
& &+\half\int_0^Ldx\,
\tilde{\Pi}_1(\tilde{\Pi}_1+2\eta).\label{Final}
\EQA
Note that, from the definition (\ref{SJinv}), the charges are not
modified, $Q_{\pm}=j^{n=0}_{\pm}=\hat{j}^{n=0}_{\pm}$.
Except for the last term that vanishes on the physical state,
(\ref{Final}) is exactly equivalent to the classical result (\ref{SIMclas}).
We may be allowed to identify the gauge invariant current $\hat{J}^\reg_1$
with $\cJ^\reg_1$ without limiting to the physical state,
\BQ
 \hat{J}^\reg_1 \Longleftrightarrow \cJ^\reg_1,
\EQ
because both of these  are invariant
 under the residual gauge transformation ($\cG_{\rm C}$).
(Classically $\hat{J}^\reg_1$ is not invariant, but it does not matter.)
This is almost trivial in the point-splitting regularization,
\BQA
J_1^\reg(x)
      &=&\lim_{\varepsilon\rightarrow 0}
	 \left\{\bpsi(x+\varepsilon)
	     \gamma_1\exp\left(ie\int_{x}^{x+\varepsilon}dyA_1(y)\right)
	 \psi(x)- {\rm v.e.v.}                               \right\}\NN
      &=&\lim_{\varepsilon\rightarrow 0}
	 \left\{\bpsi(x+\varepsilon)
	     \gamma_1\exp\left(ie\int_{x}^{x+\varepsilon}dyc\right)
	 \psi(x)- {\rm v.e.v.}                               \right\}
	-  \frac{e}{\pi}\tilde{A}_1               \NN
      &=&\cJ^\reg_1-\frac{e}{\pi}\tilde{A}_1.\label{redef}
\EQA
At this final stage,
 we can say that QHR of the Schwinger model gives the same result
 as that of the classical reduction.
While the unphysical field $\tilde{A}_1$ is absorbed into the regularized
current and the mass term for $\tilde{A}_1$ vanishes from the Hamiltonian,
the zero mode $c$ survives after the redefinition of
the currents (see Eq.\ (\ref{zeroH})).
This is because in (\ref{redef}) the zero mode should be retained
as the line integral so that $\cJ^\reg_1$ should be invariant
under the residual gauge transformation.

Since a further study of the Schwinger model lies outside the scope
of this paper and it has already been investigated intensively,
 we restrict ourselves to comment on the following two points
[\cite{IsoMura}]. First we can show that, by using the Bogoliubov
transformation, the Hamiltonian (\ref{Final}) is equivalent
to that of the massive boson with mass $m=e/\sqrt{\pi}$.
This corresponds to the bosonization of the Schwinger model.
Next if we require the invariance of the vacuum
under the large gauge transformation,
we obtain the non-trivial vacuum, the $\theta$-vacuum
that is also present in the four-dimensional SU($N$) Yang-Mills theories.


\section{SUMMARY AND DISCUSSION}

We have investigated a quantum Hamiltonian reduction based on the unitary
transformation method.  We have seen in the Schwinger model that
the regularization gives decisive effects on the unitary
transformation, but finally we recovered those in the classical
reduction.
This was achieved by a careful treatment of the regularized
currents. Note that the equivalence could not be shown  until the
Hamiltonian was written in terms of the currents.

Similar observation is expected to apply to other gauge theories. To
define a quantum system in QFT, we have to specify a regularization
 in addition to a Lagrangian.
If the system has a gauge symmetry,
the regularization should be chosen  to preserve it.
Therefore in the canonical temporal-gauge formulation,
a quantum gauge system is specified uniquely
by a set ($H,G,\cR[G]$), i.e., a Hamiltonian,
constraints which induce gauge transformations,
and a regularization that respects the gauge symmetries.
The quantum reduction is implemented by some unitary
transformations. Since the fields  before and after unitary
transformations are generally  treated without distinction, the
regularization of the operators  after reduction is the same as that
before reduction. On the other hand, the quantum reduction is
intended to simplify  the constraints, i.e., the generators of the
gauge transformations. In other words, by the reduction
transformation,  the gauge transformations inevitably change into
the ones  that are not respected by the regularization.  This
discrepancy causes a classically gauge-{\it variant} Hamiltonian
such as $H_{\rm C}$ (see Fig. 1).  Now the quantum gauge system is
unitary transformed into a set ($H',G',\cR[G]$), where $G'$ represents
residual constraints concerning the global gauge  transformations,
e.g., $Q$ in the Schwinger model on a circle.
We should note that the equivalent quantum theory to the original set
($H,G,\cR[G]$) is not ($H_{\rm cl},G',\cR[G']$) but ($H',G',\cR[G]$).
The set ($H_{\rm cl},G',\cR[G']$) is a result of the classical reduction
followed by the regularization preserving the gauge symmetries concerning $G'$.
To fill the gap between these two
results, we introduce new composite operators which can be
identified with those of the classical reduction. In the Schwinger
model, because the current construction  (the Sugawara construction)
of the Hamiltonian can be done explicitly,  the redefinition
procedure was completed only for the currents. However generally in
higher dimensional models such as the four dimensional QCD, the redefinition of
other composite operators will be needed for the proof of  their equivalence.

In non-Abelian gauge theories, we are confronted with another problem
associated with the gauge-fixing ambiguities.
We know that the Coulomb or Lorentz gauge is not a good gauge
because of the Gribov ambiguities [\cite{Gribov}].
In the presence of these ambiguities, we cannot construct gauge
invariant variables in the strictest sense.
This situation is similar to those of the Schwinger model on a circle
because it is subject to a local redundant gauge transformation,
i.e., a {\it large} gauge transformation.
As commented previously, elimination of this redundancy causes a
non-trivial structure of the vacuum.
In the recent papers on the Yang-Mills theories in a cylindrical spacetime,
it has been reported that the Gribov ambiguities also give rise to such a
non-trivial structure [\cite{HetHoso2}-\cite{LanSem}].
This effect should be investigated in higher dimensional models.

Lastly in relation to Fig. 1,
we would like to comment on the chiral Schwinger model
[\cite{Fad},\cite{JacRaj}]. It is known  as one of the anomalous
gauge theories,  which we cannot regularize while preserving gauge
symmetry.  In this paper, we have implicitly assumed that there
always exists a regularization  which respects gauge symmetry.
The scheme in Fig. 1 is limited to such theories.  In the case of
the chiral Schwinger model, it is evident that the results of the
classical and quantum Hamiltonian reductions disagree. In the
classical reduction, we can reduce the system as a gauge theory,
but in the quantum reduction, it becomes second class due to an
anomalous term in the Gauss' law  commutator [\cite{Fad}].  Now
that it is a second-class system,  the canonical temporal-gauge
formulation  cannot be applied to it. However, if we further make
an appropriate unitary transformation,  we may be able to reach a
similar Hamiltonian  to that of the classical reduction.
It might be worth while to see the differences between them.

\begin{center}{\bf ACKNOWLEDGEMENT}\end{center}
One of the authors (K.I.) would like to thank
T. Ikehashi for various discussions.

\appendix
\section{}

In this Appendix, we shall derive the fermion current algebra (\ref{Sterm1})
and the commutator (\ref{Spij}) in the text within the framework
of the $\zeta$-function regularization.
In Refs.\ [\cite{Manton},\cite{IsoMura}], the authors derived
(\ref{Sterm1}) without resorting to any regularization scheme. We
shall prove this explicitly for the regularized currents on the
fermion Fock
 space (see Ref.\ [\cite{Yazaki}] for the derivation of
(\ref{Sterm1}) by the heat-kernel method).  For simplicity, we assume
$n,m>0$.
 Using the definitions (\ref{regcurrentplus}) and (\ref{regcurrentminus}),
we have
\BQA
[(j_+^n)^\reg,&&\!\!\!\!\!\!\!\!(j_+^m)^{\rm reg \dagg}]\ket{{\rm
vac};N_+}_+\NN
   && = \lim_{s\rightarrow 0 }
       \sum_{q\in \Z} a_{q+m}^{\dagg}a_{q+n}\ket{{\rm vac};N_+}_+
	  \left\{\frac{1}{ (e_{q+m})^s (e_{q+n})^s }
		       - \frac{1}{ (e_{q})^s (e_{q})^s } \right\},
							       \label{Com1}
\EQA
where $e_q= q+\frac{\varphi}{2\pi}+\frac{ecL}{2\pi}$.
The surviving terms in the sum are classified into two cases:
\[
{\rm case}\  1\ \ \ \left \{
		       \begin{array}{l}
			  q+n\leq N_+-1\\
			  q+m\geq N_+
		       \end{array}   \right., \qquad
{\rm case}\  2\ \ \ \left \{
		       \begin{array}{l}
			  q+n\leq N_+-1\\
			  q+m = q+n
		       \end{array}   \right..
\]
In case 1, the integer $q$ is delimited to a finite number from
$N_+-m$ to $N_+-1-n$.
 (We assume $m-1\geq n$,
because otherwise there would be no contribution.)
On the other hand, in case 2, an {\it infinite} number of $q$ are involved
($q\leq N_+-1-n$).
Combining these two contributions, we can estimate Eq.\ (\ref{Com1}) as
\BQA
\lefteqn{
[(j_+^n)^\reg,(j_+^m)^{\rm reg \dagg}]\ket{{\rm vac};N_+}_+
}\NN
&=&\ \lim_{s\rightarrow 0 }\sum_{q=N_+-m}^{N_+-1-n}a_{q+m}^{\dagg}a_{q+n}
\ket{{\rm vac};N_+}_+\left\{\frac{1}{ (e_{q+m})^s (e_{q+n})^s } - \frac{1}{
(e_{q})^s (e_{q})^s } \right\} \NN
& & +\lim_{s\rightarrow 0 }\sum_{q=-\infty}^{N_+-1-n}\delta_{nm}\ket{{\rm
vac};N_+}_+
\left\{ \frac{1}{ (e_{q+n})^s (e_{q+n})^s } - \frac{1}{ (e_{q})^s (e_{q})^s }
\right\}.\label{Com2}
\EQA
 Since the first term is a finite series,
we can take the $s\rightarrow 0$ limit first.
Then this term vanishes.
Though we cannot change the order of the limit
and the summation in the second term
of (\ref{Com2}), we can estimate it using the $\zeta$-function:
\BQA
\lefteqn{
\lim_{s\rightarrow 0}\sum_{q=-\infty}^{N_+-1-n}
       \left\{ \frac{1}{(e_{q+n})^s(e_{q+n})^s}- \frac{1}{(e_q)^s(e_q)^s}
       \right\} }\NN
&=&\lim_{s\rightarrow 0}\sum_{q=-\infty}^{N_+-1-n}
	 \left\{ \frac{1}{(q+n+\frac{\varphi}{2\pi}+\frac{ecL}{2\pi})^{2s}}-
		     \frac{1}{(q+\frac{\varphi}{2\pi}+\frac{ecL}{2\pi})^{2s}}   \right\}\NN
&=&\lim_{s\rightarrow 0}
	\left\{ \zeta\left(2s,1-\frac{\varphi}{2\pi}-\frac{ecL}{2\pi}-N_+\right)
	      - \zeta\left(2s,1-\frac{\varphi}{2\pi}-\frac{ecL}{2\pi}-N_++n\right)
\right\}\NN
&=&n.
\EQA
Now we obtain Eq.\ (\ref{Sterm1}) on the $N_+$-vacuum.
We can also evaluate the summation of infinite series
without using the $\zeta$-function regularization:
\BQA
\lim_{s\rightarrow 0}\sum_{q=-\infty}^{N_+-1-n}
   \left\{ \frac{1}{(e_{q+n})^s(e_{q+n})^s}- \frac{1}{(e_q)^s(e_q)^s}  \right\}
&=&\lim_{s\rightarrow 0} \left\{ \sum_{q=-\infty}^{N_+-1}
		 - \sum_{q=-\infty}^{N_+-1-n} \right\} \frac{1}{e_q^{2s}} \NN
&=&\lim_{s\rightarrow 0}\sum_{q=N_+-2-n}^{N_+-1} \frac{1}{e_q^{2s}} \NN
&=&\sum_{q=N_+-2-n}^{N_+-1} 1\  =n,
\EQA
where we have changed the order, because the infinite series can be made
finite.
 The authors of Refs.\ [\cite{Manton},\cite{IsoMura}] were able to
derive Eq.\ (\ref{Sterm1})
 without resorting to any regularization
 because they proceeded in such a way
that the number of non-vanishing terms is always finite.
So far as the number of creation and annihilation operators
acting on the $N_+$-vacuum is finite, we can follow the same way
and obtain the same results for arbitrary
states.  It is straightforward to extend the calculation to the
negative-chirality sector.

It is not trivial to derive the commutator (\ref{Spij})
by the $\zeta$-function regularization.
To begin with, we have to calculate the commutator between the
annihilation operator $a_n$ and $\Pi_1$.  The basis function
$\chi_n$, a functional of $c$, does not commute with
$\Pi_1$,
\BQ
[\,\chi_n(x), \, \Pi_1(y)\,
]=-e\left(\tilde{\theta}(x-y)+\tilde{\theta}(y)\right)\chi_n(x),
\EQ
but since $[\psi(x),\Pi_1(y)\,]=0$, we obtain the commutator,
\BQ
[\, a_n,\, \Pi_1(y)\, ]=ea_n\tilde{\theta}(y)+
\sum_{q\neq 0}a_{n+q}\frac{e}{2\pi iq}\ \e^{-\frac{2\pi iq}{L}y}.
\EQ
After some manipulations, we have
\BQ
[\, \Pi_1(x),\, (j_+^n)^\reg\, ]
=-\lim_{s\rightarrow 0}\sum_{p\in \Z}\sum_{q\neq 0}a_p^{\dagg}a_{p+q+n}
     \left\{\frac{1}{(\lambda \varepsilon_p)^s}-\frac{1}{(\lambda
\varepsilon_{p+q})^s}\right\}\frac{e}{2\pi iq}\ \e^{-\frac{2\pi iq}{L}x}.
\EQ
If this is estimated on the $N_+$-vacuum,
the surviving terms are classified into a finite series and an
infinite series.  Since the finite series does not contribute,
the commutator reduces to
\BQA
\lefteqn{[\Pi_1(x),(j_+^n)^\reg]\ket{{\rm vac};N_+}_+}\NN
&=&-\lim_{s\rightarrow 0}\sum_{p=-\infty}^{N_+-1}\ket{{\rm vac};N_+}_+
\left\{\frac{1}{(\lambda \varepsilon_p)^s}
-\frac{1}{(\lambda \varepsilon_{p-n})^s}\right\}\frac{e}{-2\pi in}\
\e^{\frac{2\pi in}{L}x}\quad (n\neq 0)\NN
&=&\frac{e}{2\pi i}\ \e^{\frac{2\pi i n}{L}x}\ket{{\rm vac};N_+}_+
\quad (n\neq 0).
\EQA
When $n=0$, we have $[\,\Pi_1(x),(j^{n=0}_+)^\reg\,]=0$.
This leads to
\BQ
[\,\Pi_1(x),J_+^\reg(y)\,]\ket{{\rm vac};N_+}_+
=\frac{e}{2\pi i}\left(\delta(x-y)-\frac{1}{L} \right)\ket{{\rm vac};N_+}_+.
\EQ
Combining this and the similar results for the negative-chirality
sector, we finally obtain
\BQ
[\,\Pi_1(x),J^\reg_1(y)\,]\ket{{\rm vac};N_+,N_-}
=-\frac{ie}{\pi}\left(\delta(x-y)-\frac{1}{L} \right)\ket{{\rm vac};N_+,N_-}.
\EQ
The same relation holds for general states as long as
the number of creation or annihilation operators
acting on the vacuum is finite.

\begin{figurecaptions}
\item{} Starting from a classical unreduced system whose constraint
is $G=0$, the upper approach corresponds to a classical
reduction, and the lower approach to a quantum reduction.
In the classical reduction, the constraint becomes $G'=0$ after the
reduction of the phase space, and regularization  ($\cR[G']$) is
performed to preserve the gauge symmetry generated by $G'$. On the
other hand, in the quantum reduction, the regularization ($\cR[G]$)
respects the gauge symmetry generated by $G$. As in the Schwinger
model, these two results are expected to be equivalent by some
redefinitions of composite operators such as currents.
\end{figurecaptions}

{
\begin{figure}[ht]
\setlength{\unitlength}{1mm}
\begin{picture}(150,70)(-10,40)
      \put(0,70){\framebox(30,16){\shortstack{$H$ \\$G=0$}}}
     \put(40,90){\framebox(30,16){\shortstack{$H_{\rm cl} $\\$G'=0$}}}
     \put(40,50){\framebox(30,16){\shortstack{
				     $H'$, $\cR[G]$ \\$G\ketrm{phys}\!=\!0$}}}
     \put(100,90){\framebox(30,16){\shortstack{
				     $(H_{\rm cl})^\reg$, $\cR[G']$\\$G'\ketrm{phys}\!=\!0$}}}
     \put(100,50){\framebox(30,16){\shortstack{
				     $H'$, $\cR[G]$\\$G'\ketrm{phys}'\!=\!0$}}}

     \put(70,58){\vector(1,0){30}}
     \put(70,98){\vector(1,0){30}}
     \put(15,98){\vector(1,0){25}}
     \put(15,58){\vector(1,0){25}}
     \put(15,86){\line(0,1){12}}
     \put(15,58){\line(0,1){12}}
     \put(115,66){\vector(0,1){24}}

      \put(20,100){reduction}
      \put(16,54){quantization}
      \put(74,100){quantization}
      \put(72,54){\shortstack{unitary\\transformation}}
      \put(116,78){\shortstack{some\\redefinitions}}
    \end{picture}
\end{figure}
}

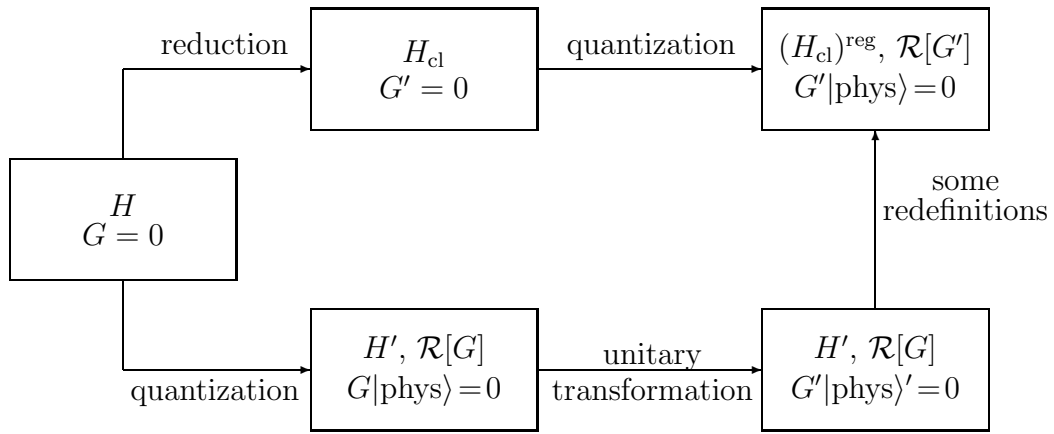
\end{document}